\documentclass[letterpaper]{article} 
\usepackage[]{aaai25}  
\usepackage{aaai25}
\usepackage{abstract}
\usepackage{times}  
\usepackage{helvet}  
\usepackage{courier}  
\usepackage[hyphens]{url}  
\usepackage{graphicx} 
\usepackage{amsmath}
\usepackage{amssymb}
\usepackage{subcaption}
\usepackage{booktabs}
\usepackage{subfloat}

\usepackage{natbib}  
\usepackage{caption} 
\usepackage{amsmath}
\usepackage{amssymb}
\usepackage{algpseudocode}
\usepackage{algorithmicx,algorithm}
\usepackage{color}
\definecolor{S1}{RGB}{45,86,174}
\definecolor{S11}{RGB}{42,80,162}
\definecolor{S12}{RGB}{29,45,112}
\definecolor{S13}{RGB}{20,71,83}
\definecolor{S2}{RGB}{41,22,140}
\definecolor{S3}{RGB}{0,124,130}

\usepackage{newfloat}
\usepackage{listings}
\DeclareCaptionStyle{ruled}{labelfont=normalfont,labelsep=colon,strut=off} 
\floatstyle{ruled}
\newfloat{listing}{tb}{lst}{}
\floatname{listing}{Listing}
\makeatletter

\newcommand{\Rmnum}[1]{\expandafter\@slowromancap\romannumeral #1@}
\makeatother

\title{AoI-MDP: An AoI Optimized Markov Decision Process\\(Student Abstract)}
\author{
    Yimian Ding\textsuperscript{\rm 1},  
    Jingzehua Xu\textsuperscript{\rm 1},
    Yiyuan Yang\textsuperscript{\rm 2},
    Guanwen Xie\textsuperscript{\rm 1},
    Xinqi Wang\textsuperscript{\rm 1}\equalcontrib,
    Shuai Zhang\textsuperscript{\rm 3}\equalcontrib
}
\affiliations{
    \textsuperscript{\rm 1}Tsinghua Shenzhen International Graduate School, Tsinghua University, China\\
    \textsuperscript{\rm 2}Department of Computer Science, University of Oxford, United Kingdom\\
    \textsuperscript{\rm 3}Department of Data Science, New Jersey Institute of Technology, USA\\

    yimiandingthu@gmail.com, xjzh23@mails.tsinghua.edu.cn,
    yiyuan.yang@cs.ox.ac.uk,
    gwxie360@outlook.com,
    aaahayooha@163.com,
    sz457@njit.edu

}

\begin{document}

\maketitle

%

\section{Abstract}
Ocean exploration places high demands on autonomous underwater vehicles, especially when there's observation delay. We propose age of information optimized Markov decision process (AoI-MDP) to enhance underwater tasks by modeling observation delay as signal delay and including it in the state space. AoI-MDP also introduces wait time in the action space and integrates AoI with reward functions, optimizing information freshness and decision-making using reinforcement learning. Simulations show AoI-MDP outperforms the standard MDP, demonstrating superior performance, feasibility, and generalization in underwater tasks. To accelerate relevant research, we have made the codes available as open-source at https://github.com/Xiboxtg/AoI-MDP.

\section{Introduction}\label{se:1}
Utilizing autonomous underwater vehicles (AUVs) with reinforcement learning (RL) is a significant research focus for ocean exploration \cite{1,2,5}. However, tasks often fail due to observation delays caused by information limitation, leading to non-causality in control policies \cite{ding2024multi,4, yi2020deep}. Most studies use the standard Markov decision process (MDP) without accounting for observation delay, assuming instant state information reception\cite{10650551,6}. This idealization is impractical due to information delay effects and channel limitation, which reduce information freshness and decision-making efficiency \cite{10.1145/149439.133106,7}.

Based on above analysis, we propose an age of information (AoI) optimized MDP (AoI-MDP) for underwater tasks to improve performance with observation delay. Our contributions include the following:
\begin{itemize}
\item To the best of our knowledge, we are the first to model underwater tasks as an MDP with observation delay and AoI, using RL for AUV training to optimize information updating and decision-making strategies jointly.

\item We utilize statistical delay modeling (SDM) for delay-oriented modeling of observation delay via sensor-based model, yielding realistic results.

\item Comprehensive experiments in the underwater data collection task show AoI-MDP's superior feasibility and performance in balancing multi-objective optimization.
\end{itemize}

\section{Methodology}\label{se:3}
\textbf{AoI Optimized Markov Decision Process.} As illustrated in Figure 1, consider the $i$-th delayed observation signal is transmitted from the AUV at time $T_i$, and the corresponding observed information is received at time $D_i$, AoI is defined using a sawtooth piecewise function
\begin{equation}\Delta(t)=t-T_i,D_i\leq t<D_{i+1}, \forall i \in \mathbf{N}.\end{equation}

\begin{figure}[!t]
    \centering
    \includegraphics[width=0.948\linewidth]{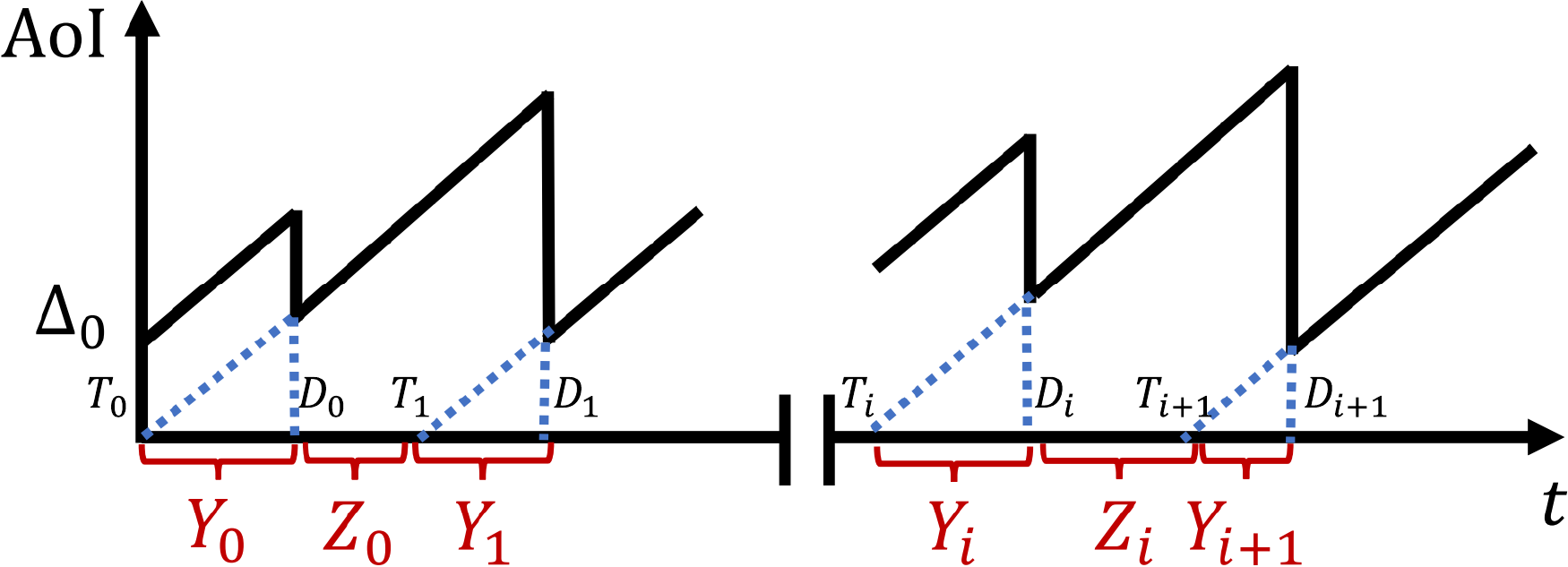}
    \caption{Illustration of the AoI model.}
    \label{fig_1}
    \end{figure}

In AoI-MDP, AoI is not only part of reward functions to guide objective optimization via RL, but also crucial side information for decision-making. The AoI-MDP's state space, action space, and reward functions are detailed as follows:

\textbf{State Space $\boldsymbol{\mathcal{S}}$:} consists of the AUV's observed information $s^{'}_i$ and observation delay $Y_i$ at time $i$, represented by $s_i=(s^{'}_i, Y_i) \in \boldsymbol{\mathcal{S^{'}}} \times \boldsymbol{Y}$. Delay-oriented modeling of $s^{'}_i$ and $Y_i$ is achieved via SDM, whose details are presented later. 

\textbf{Action Space $\boldsymbol{\mathcal{A}}$:} 
composed of the tuple $a_i=(a_i^{'}, Z_i) \in \boldsymbol{\mathcal{A^{'}}} \times \boldsymbol{Z}$, where $a_i^{'}$ denotes the actions taken by the AUV and $Z_i$ indicates the wait time between observing the environmental information and decision-making at time $i$.

\textbf{Reward Function $\boldsymbol{\mathcal{R}}$:} Apart from the original reward function $r_i^{'}$ in standard MDP, AoI-MDP introduces the time-averaged AoI as a new reward component. Thus, the updated reward function can be represented by tuple $r_i=(r^{'}_i, -\Bar{\Delta})$. And the time-averaged AoI can be computed as follows: $\Bar{\Delta}=\frac{\sum_{i=1}^{\mathcal{N}}((2Y_{i-1}+Y_i+Z_{i-1})\times(Y_i+Z_{i-1}))+S_0}{2\times(\sum_{i=1}^{\mathcal{N}}Z_{i-1}+\sum_{i=1}^{\mathcal{N}}Y_{i}+Y_0)},$
where $\mathcal{N}$ is the length of information signal, $S_0=0.5 \times (2\Delta_0+Y_0) \times Y_0$.

\!\!\!\!\!\!\textbf{Observation Delay and Information Modeling.}
Different from previous work, our study enhances the state space of AoI-MDP by incorporating observed information from AUV sensors and treating observation delay as delayed observation signal delay, as shown in Figure 2.

Specifically, our study assumes the AUV uses a sensing model to estimate distances to environmental objects. We further employ the delay estimator as an estimator to determine the time delay, which can be represented as
\begin{subequations}
\begin{align}
\mathcal{X}[n]=\mathcal{S}[n-&Y_i]+\mathcal{W}[n], n=0,1,\dots,N-1,\\
J[Y_i]=\sum_{n={Y_i}}^{Y_i+M-1}&\mathcal{X}[n]\mathcal{S}[n-Y_i], 0\leq Y_i\leq N-M,\\
&\hat{Y_i} = {\rm argmax} \left[J[Y_i] \right],
\end{align}
\end{subequations}
where $\mathcal{S}[n]$ represents the known sequence, while $Y_i$ denotes the time delay to be estimated, and $\mathcal{W}[n]$ is Gaussian white noise with variance $\sigma^2$.
where $M$ is sampling length of $\mathcal{S}[n]$.

\begin{figure}[!t]
    \centering
    \includegraphics[width=1.0\linewidth]{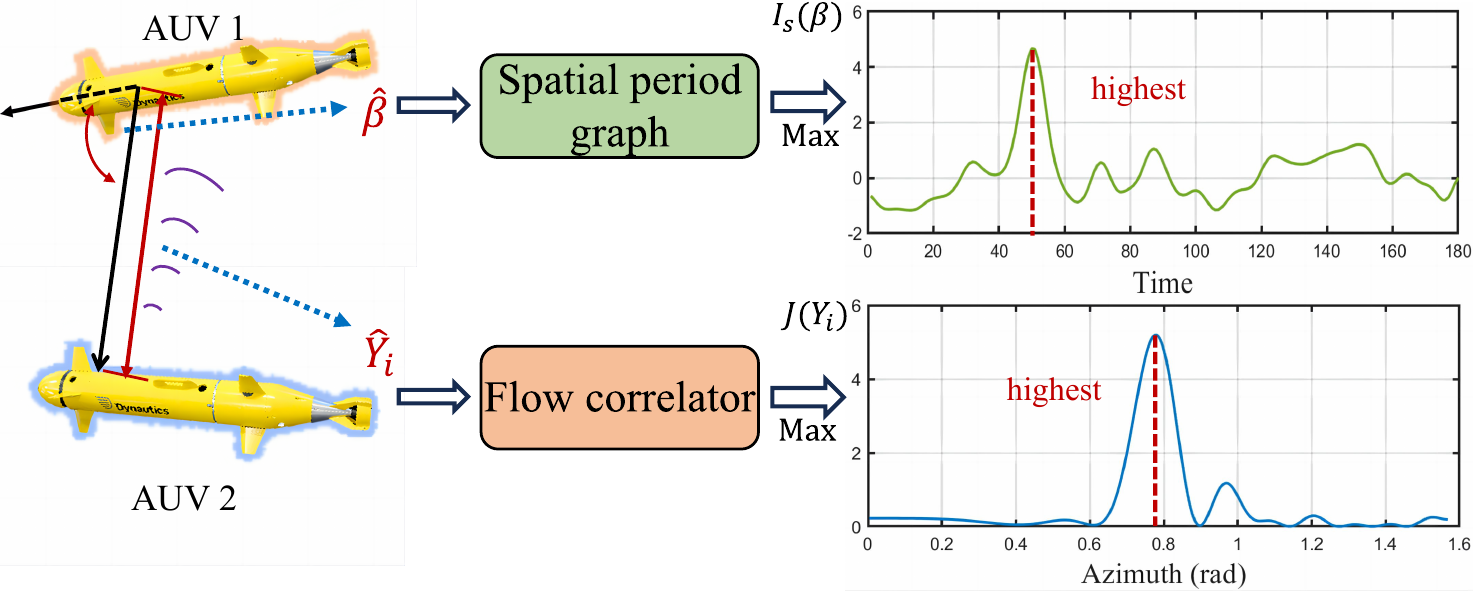}
    \caption{Diagram of the heading and time delay estimation.}
    \label{fig_2}
\end{figure}

On the other hand, the AUV in our study utilizes a generic sensing approach to estimate the heading $\beta$ between its orientation and environmental objects. The estimator in SDM is further leveraged to estimate the heading $\beta$. By maximizing the spatial period graph, the estimate of $\beta$  (0\textless$\beta$\textless$\pi$/2) can be calculated as follows:

\begin{subequations}
\begin{align}
\quad x \left[n\right] = & A c o s \left[2 \pi \left(F_{0} \frac{d}{c} c o s \beta\right) n + \phi\right] + \mathcal{W}[n] , \textrm{ } \notag\\
&\quad\quad\quad n = 0,1 , \cdot \cdot \cdot , M - 1,\\
\!\!\!\!\!\!\!\!\!I_{s} \!\left(\beta\right) \!\!=\!\! \frac{1}{M}\! &\left(\!\left|\!\!\right. \sum_{n = 0}^{M - 1} x \left[n\right] {\rm exp} \left[\right.\! -\! j 2 \pi \!\left(\right.\! F_{0} \frac{d}{c} c o s \beta\! \left.\right)\!n \left]\right. \left|\right.\right)^{2}\!\!\!\!,\\
&\quad\quad\quad\hat{\beta} = {\rm argmax} \left[I_{s} \left(\beta\right)\right],
\end{align}
\end{subequations}
where $F_0$ denotes the frequency of observation signal, while $d$ represents the interval between sensors. Besides, $c$ indicates the speed of delayed observation signal propagation.

\section{Simulation Experiments}\label{se:4}
Since open-source underwater tasks are scarce, we use a multi-AUV data collection task to evaluate the AoI-MDP's feasibility and effectiveness. For more details and parameters, refer to the previous work \cite{3}.

We compared RL training results based on AoI-MDP and standard MDP under identical conditions, respectively. As shown in Figure 3, AoI-MDP achieved lower time-averaged AoI, reduced energy consumption, higher sum info rate, and greater cumulative rewards, indicating improved training effectiveness and performance.

Then we assessed AoI-MDP's generalization using common delay models (exponential, poisson, geometric) and compared the results with the SDM model, as shown in Table 1. AoI-MDP demonstrated superior performance across various distributions, highlighting its strong generalization capabilities. The SDM model achieved near-optimal results in AoI, data rate, and energy consumption optimization, proving effective in the underwater data collection task.

\begin{figure}[!t]
    \centering
    \begin{subfigure}{0.23\textwidth}
        \centering
        \includegraphics[width=\linewidth]{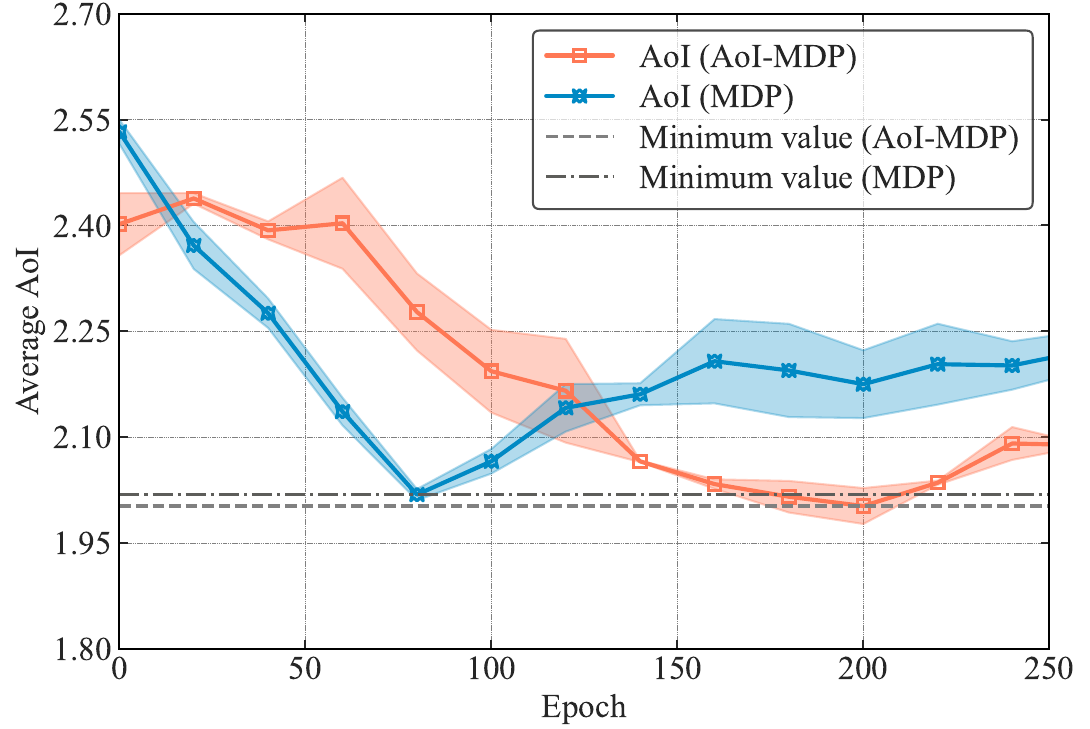}
        \caption{Average AoI}
        \label{fig:subfig1}
    \end{subfigure}
    \hfill
    \begin{subfigure}{0.23\textwidth}
        \centering
        \includegraphics[width=\linewidth]{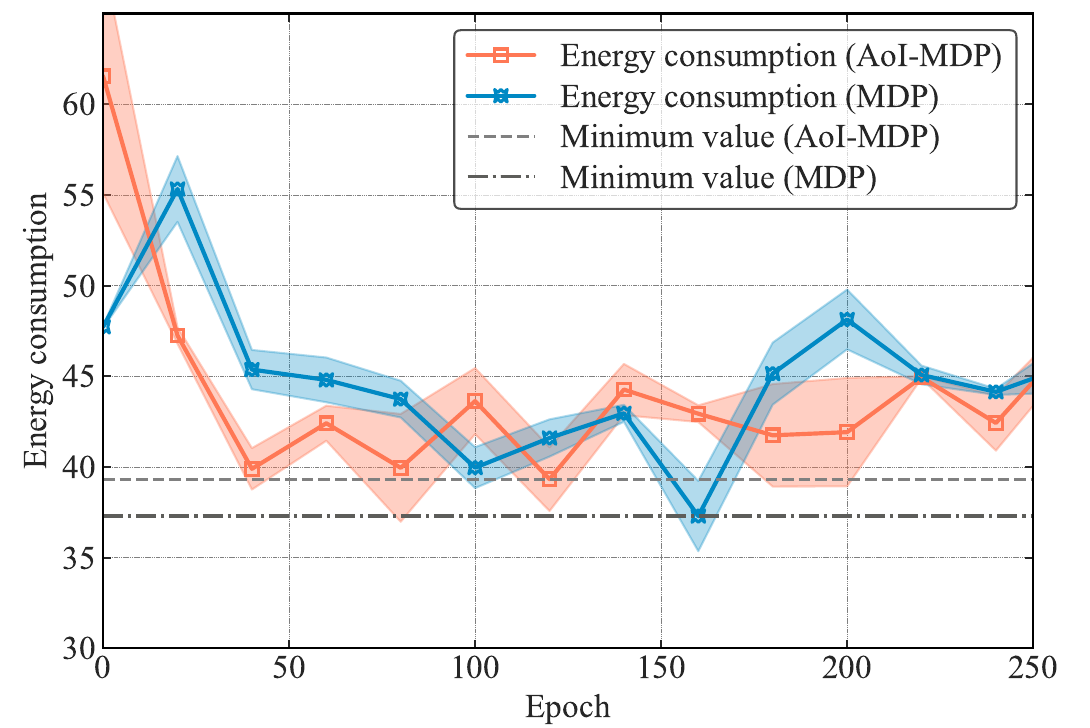}
        \caption{Energy consumption (EC)}
        \label{fig:subfig2}
    \end{subfigure}
    
    
    \begin{subfigure}{0.23\textwidth}
        \centering
        \includegraphics[width=\linewidth]{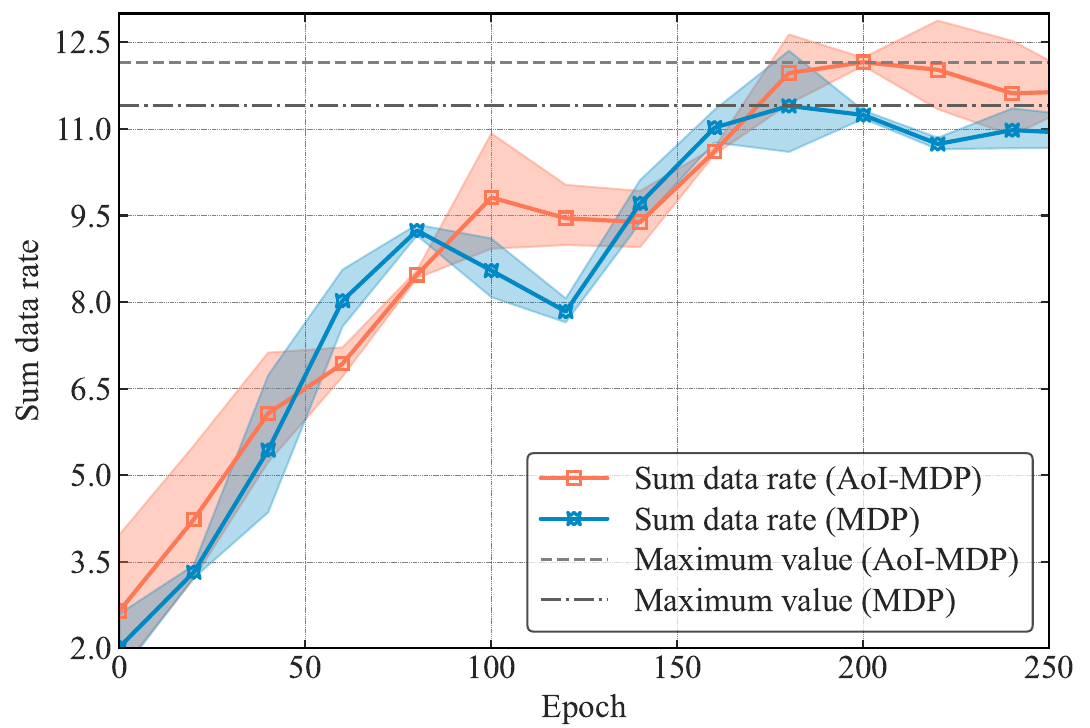}
        \caption{Sum data rate (SIR)}
        \label{fig:subfig3}
    \end{subfigure}
    \hfill
    \begin{subfigure}{0.23\textwidth}
        \centering
        \includegraphics[width=\linewidth]{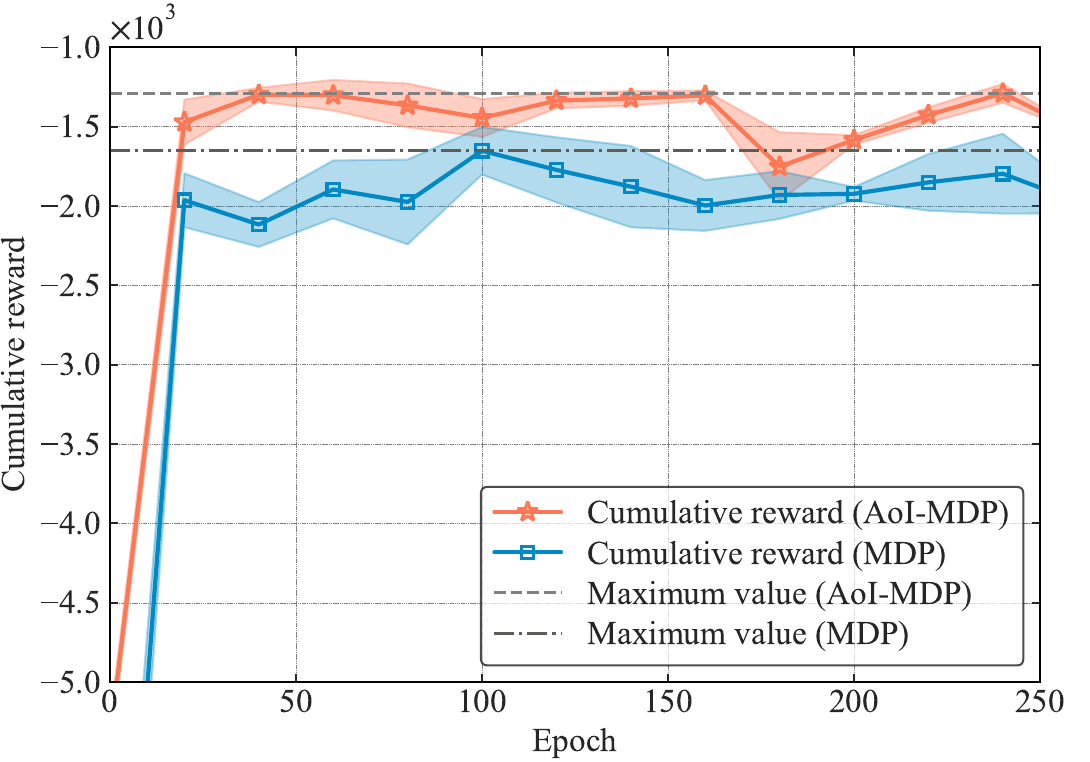}
        \caption{Cumulative reward}
        \label{fig:subfig4}
    \end{subfigure}
    
    \caption{Comparison of experimental results of RL training based on AoI-MDP and standard MDP.}
    \label{fig:overall}
\end{figure}

\begin{table}[!t]
    \centering
    \caption{Comparison of different delay models.}
    \label{table:masking_performance}
    \begin{tabular}{lcccc}
    \toprule
      & AoI & SIR & EC  \\
    \midrule
    SDM          & \textbf{1.97±0.26} & \textbf{11.99±0.73} & \textbf{33.83±2.59} \\
    Poisson      & 3.42±0.18  & 5.95±2.42  & 34.27±7.98  \\
    Exponential  & 2.67±0.26   & 7.65±1.99  & 43.11±4.16  \\
    Geometric    & 2.38±0.28  & 12.34±0.79 & 58.15±9.49 \\
    \bottomrule
    \end{tabular}
\end{table}

Finally, given the limited space, we have made the codes and supplementary materials available as open-source at our arXiv version:
https://arxiv.org/abs/2409.02424.

\section{Conclusion}

In this work, we introduce the AoI-MDP into the underwater task. By integrating wait time into the action space and linking AoI with the reward function, AoI-MDP enhances both information freshness and decision-making processes through reinforcement learning. Simulations confirm the AoI-MDP’s superior performance in underwater tasks. 

\section{Acknowledgment}

Part of this work was done when Yimian Ding and Jingzehua Xu were studying in the MicroMasters Program in Statistics and Data Science at Massachusetts Institute of Technology (MIT). We are very grateful to Yiyuan Yang and Shuai Zhang at University of Oxford and New Jersey Institute of Technology (NJIT) for their strong support, and to Miao Liu and Songtao Lu at IBM research and MIT-IBM Watson AI Lab and for their valuable advice, respectively. Additionally, we thank all anonymous reviewers for their constructive comments.

\section{Impact Statement}

This paper presents work whose goal is to explore using AoI-MDP to enhance the performance of marine robotics for the underwater tasks. There are many potential societal consequences of our work, none of which we feel must be specifically highlighted here.

\bibliography{aaai25}

\end{document}